\pdfoutput=1
\documentclass[11pt]{article}

\usepackage[margin=1in]{geometry}
\usepackage[T1]{fontenc}
\usepackage[utf8]{inputenc}

\usepackage{amsmath,amssymb,amsfonts,amsthm,mathtools}
\usepackage{braket}
\usepackage{microtype}
\usepackage{booktabs}
\usepackage{enumitem}
\usepackage{xcolor}
\usepackage{xspace}

\usepackage{graphicx}
\usepackage{float}
\usepackage{hyperref}
\usepackage[nameinlink]{cleveref}

\usepackage{tikz}
\usepackage{quantikz}
\usetikzlibrary{arrows.meta,positioning,shapes.geometric}

\hypersetup{colorlinks=true,linkcolor=blue,citecolor=blue,urlcolor=blue}

\newtheorem{theorem}{Theorem}[section]
\newtheorem{lemma}[theorem]{Lemma}
\newtheorem{proposition}[theorem]{Proposition}
\newtheorem{corollary}[theorem]{Corollary}
\newtheorem{definition}{Definition}[section]
\newtheorem{remark}{Remark}[section]

\newcommand{\qRAM}{QRAM\xspace}
\newcommand{\QRAM}{\textnormal{\textsc{QRAM}}\xspace}
\newcommand{\Uqram}{U_{\mathrm{QRAM}}}
\newcommand{\Id}{\mathbb{I}}
\renewcommand{\proj}[1]{\ket{#1}\!\bra{#1}}

\newcommand{\Pone}[1]{\proj{1}_{#1}}
\newcommand{\Paddr}[1]{\proj{#1}_{A}}

\newcommand{\Htot}{H_{\mathrm{tot}}}
\newcommand{\Hcell}{H_{a,j}}
\newcommand{\Hcellone}{H^{(1)}_{a,j}}

\title{\vspace{-0.4cm}
Universal Quantum Random Access Memory:\\
A Data-Independent Unitary with a Commuting-Projector Hamiltonian\\[0.3em]
\large Mode-Addressed 3-Local Cells and Explicit Latency Assumptions
\vspace{-0.25cm}
}
\author{Leonardo Bohac}
\date{January 2026}

\begin{document}
\maketitle
\vspace{-0.35cm}

% ==============================================================================
\begin{abstract}
Quantum random access memory (\qRAM) is a central primitive for coherent data access in quantum algorithms, yet it remains controversial in practice because the wall-clock cost of ``one lookup'' can hide routing depth, control overhead, and geometric constraints.
We present a universal \QRAM construction (U-\QRAM) in which the database is a physical memory register that participates in the lookup unitary as quantum control.
This yields a single fixed, data-independent lookup unitary on $A\otimes D\otimes M$ that is correct for all basis-encoded databases.

Our first contribution is an explicit, exact Hamiltonian realization: U-\QRAM equals a single time-independent evolution $\Uqram=\exp(-i\Htot)$ where $\Htot$ is a sum of mutually commuting projector terms, one per memory cell, and the construction avoids control-dependent phase ambiguities.
Our second contribution is an architectural sharpening: under unary (one-hot, mode-addressed) encoding of the address, each cell term becomes uniform and 3-local, delineating the most direct path toward constant-latency interpretations.
We keep claims conservative by separating \emph{latency} from \emph{work} and stating explicit hardware assumptions required for constant wall-clock queries.
\end{abstract}

\vspace{-0.1cm}
\begin{center}
\textbf{Contributions.}
\end{center}
\vspace{-0.2cm}
\begin{enumerate}[leftmargin=1.15cm, itemsep=2pt]
\item \textbf{Universal, data-independent lookup unitary with the database in Hilbert space.}
We give a canonical projector-form definition of U-\QRAM as a single unitary acting on $A\otimes D\otimes M$.
\item \textbf{Exact commuting-projector Hamiltonian form.}
We show $\Uqram=\exp(-i\Htot)$ for an explicit time-independent $\Htot=\sum_{a,j}\Hcell$ with $[\Hcell, H_{b,\ell}]=0$.
\item \textbf{Mode-addressed reduction to 3-local cell interactions.}
With unary/one-hot addressing, each $\Hcell$ reduces to a uniform 3-local form coupling only an address mode, a memory qubit, and an output qubit.
\item \textbf{Latency regimes and explicit assumptions.}
We separate latency/work/parallelism and state explicit assumptions under which ``$O(1)$ query time'' becomes physically meaningful.
\end{enumerate}

% ==============================================================================
\section{Introduction: what is disputed about \qRAM}
% ==============================================================================

The canonical \qRAM specification is
\begin{equation}
\ket{a}_A\ket{y}_D\ket{\mathrm{mem}}_M\ \longmapsto\ \ket{a}_A\ket{y\oplus x_a}_D\ket{\mathrm{mem}}_M,
\label{eq:qram-spec}
\end{equation}
where $A$ is an address register, $D$ is an output (data) register, $M$ is a memory register storing $N$ words of $K$ bits, and $x_a\in\{0,1\}^K$ denotes the word at address $a$.
The controversy is not about \cref{eq:qram-spec}, but about the \emph{cost model} behind ``one coherent query'': routing depth, signal propagation, control complexity, and fault-tolerant overhead can wash out oracle-model speedups \cite{aaronson2015fineprint,jaquesRattew2025survey,wang2024causalbounds}.

\paragraph{What this paper does.}
We present a universal \qRAM construction (U-\QRAM) in which the database is a physical register that participates in the unitary as quantum control.
This yields a single fixed lookup unitary on $A\otimes D\otimes M$ that is correct for \emph{all} basis-encoded databases.
We then provide an explicit, exact Hamiltonian $\Htot$ such that a single time-independent evolution $\exp(-i\Htot)$ realizes the lookup, and we show that unary (one-hot, mode-addressed) addressing reduces each cell interaction to 3-local form.

\paragraph{What this paper does \emph{not} claim.}
We do not claim that commutation alone guarantees constant wall-clock time.
Constant latency is meaningful only under explicit assumptions about global coupling, parallel hardware, and geometry, and remains subject to causal/locality bounds \cite{wang2024causalbounds}.
We also do not claim that U-\QRAM removes the need for error correction in fault-tolerant settings.
Our goal is to replace vague oracle assumptions by a concrete target: a fixed unitary together with an explicit Hamiltonian and stated physical assumptions.

\begin{remark}[Conventions]
We work in units $\hbar=1$.
Writing $\Uqram=\exp(-i\Htot)$ absorbs the evolution time into $\Htot$.
Equivalently, one may write $\Uqram=\exp(-i\tau H_{\mathrm{phys}})$ for a fixed physical time $\tau$.
\end{remark}

% ==============================================================================
\section{Model: registers and the U-\QRAM lookup unitary}
\label{sec:model}
% ==============================================================================

Let $N=2^n$ be the number of addresses and $K$ the word size.
We use three registers:
\begin{itemize}[nosep]
\item \textbf{Address register} $A$:
  either \textbf{binary} ($n=\log_2 N$ qubits, basis $\{\ket{a}\}_{a=0}^{N-1}$) or \textbf{unary/one-hot} (\cref{sec:mode}).
\item \textbf{Output register} $D$ with $K$ qubits $d_0,\dots,d_{K-1}$, basis $\ket{y}$.
\item \textbf{Memory register} $M$ with $NK$ qubits $m_{a,j}$, one per cell $(a,j)$.
\end{itemize}

For a basis-encoded (classical) database, we write
\begin{equation}
\ket{\mathrm{mem}}_M
=\bigotimes_{a=0}^{N-1}\bigotimes_{j=0}^{K-1}\ket{x_{a,j}}_{m_{a,j}},
\qquad x_{a,j}\in\{0,1\},
\label{eq:mem-basis}
\end{equation}
where $x_{a,\cdot}\in\{0,1\}^K$ is the word at address $a$.

\subsection{A canonical projector-form definition}

Define address projectors $P_a:=\Paddr{a}$ and memory-bit projectors $Q_{a,j}:=\Pone{m_{a,j}}$.
Let $X_{d_j}$ be Pauli-$X$ on output qubit $d_j$ (identity elsewhere).

\begin{definition}[U-\QRAM lookup unitary]
\label{def:uqram}
The U-\QRAM lookup unitary is
\begin{equation}
\Uqram
:= \prod_{a=0}^{N-1}\prod_{j=0}^{K-1}
\Bigl( (\Id - P_a Q_{a,j})\otimes \Id_D \, +\, (P_a Q_{a,j})\otimes X_{d_j}\Bigr),
\label{eq:uqram-product}
\end{equation}
acting on $A\otimes D\otimes M$.
Equivalently: for each cell $(a,j)$, ``flip $d_j$ iff $A=a$ and $m_{a,j}=1$.''
\end{definition}

\begin{proposition}[Correctness for basis-encoded databases]
\label{prop:correctness}
If $M$ is prepared as in \cref{eq:mem-basis}, then for every basis address $\ket{a}_A$ and output basis state $\ket{y}_D$,
\begin{equation}
\Uqram\,\ket{a}_A\ket{y}_D\ket{\mathrm{mem}}_M
= \ket{a}_A\ket{y\oplus x_a}_D\ket{\mathrm{mem}}_M.
\end{equation}
\end{proposition}

\begin{proof}
Fix $a$ and $j$.
The factor indexed by $(a,j)$ applies $X_{d_j}$ exactly on the subspace where $A=a$ and $m_{a,j}=1$, and applies identity otherwise.
Across all $j$, the output bits are XORed with the stored word $x_a$.
\end{proof}

\begin{remark}[If the memory is quantum]
If $M$ is placed in a superposition over computational-basis strings, \cref{eq:uqram-product} is still a single well-defined unitary, but it generally \emph{entangles} $D$ with $M$.
The classical database behavior \cref{eq:qram-spec} is recovered when $M$ is prepared in a computational-basis encoding of classical data.
\end{remark}

\subsection{Block-diagonal viewpoint (memory as a configuration register)}
\label{sec:block}

Because $\Uqram$ is composed of controlled-$X$ gates, it is a permutation in the joint computational basis.
It is also natural to view $M$ as a \emph{configuration register}: conditioned on each computational-basis state of $M$, the operator on $A\otimes D$ is a fixed address-controlled XOR.
Formally, expanding in the memory basis gives
\begin{equation}
\Uqram
= \bigoplus_{\mathrm{mem}\in\{0,1\}^{NK}} U_{\mathrm{mem}},
\qquad
U_{\mathrm{mem}}\in \mathbb{C}^{(N2^K)\times(N2^K)},
\end{equation}
where each $U_{\mathrm{mem}}$ acts only on $A\otimes D$ and is itself block-diagonal over addresses:
\begin{equation}
U_{\mathrm{mem}} = \bigoplus_{a=0}^{N-1} \Bigl( \bigotimes_{j=0}^{K-1} X^{x_{a,j}} \Bigr).
\label{eq:block-form}
\end{equation}
This makes explicit that U-\QRAM is the canonical ``address-select, then XOR'' permutation, with the database string determining which $X$ factors occur in each address block.

\subsection{Permutation and involution structure}

\begin{proposition}[Permutation unitary]
\label{prop:perm}
$\Uqram$ is a permutation matrix in the joint computational basis of $A\otimes D\otimes M$.
\end{proposition}

\begin{proof}
Each factor in \cref{eq:uqram-product} is a controlled-$X$ gate (possibly with multi-control $P_aQ_{a,j}$) and therefore permutes computational basis states.
A product of permutations is a permutation.
\end{proof}

\begin{lemma}[Commutation of cell factors]
\label{lem:factors-commute}
All factors in \cref{eq:uqram-product} commute.
\end{lemma}

\begin{proof}
For distinct addresses $a\neq b$, $P_aP_b=0$ and $P_a$ commutes with $P_b$.
All $Q_{a,j}$ commute since they are projectors on distinct memory qubits.
For distinct output bits $j\neq \ell$, $X_{d_j}$ commutes with $X_{d_\ell}$.
Thus the controlled-$X$ factors commute.
\end{proof}

\begin{corollary}[Self-inverse]
\label{cor:selfinverse}
$\Uqram^\dagger=\Uqram^{-1}=\Uqram$.
\end{corollary}

\begin{proof}
Each factor in \cref{eq:uqram-product} is Hermitian and involutory (controlled-$X$ squares to identity) and, by \cref{lem:factors-commute}, the product inherits these properties.
\end{proof}

\begin{remark}[On ``uniqueness'']
The lookup specification \cref{eq:qram-spec} constrains the action of $\Uqram$ only on basis states where $M$ is unchanged and the XOR rule holds for all $y\in\{0,1\}^K$.
Within the class of permutation unitaries generated by controlled-$X$ factors of the form in \cref{eq:uqram-product}, these constraints fix the construction essentially uniquely.
More generally, if one relaxes the class of allowed unitaries (e.g., allowing arbitrary phases on subspaces not probed by a restricted test set), additional degrees of freedom can appear.
This is one reason to keep hardware claims tied to explicit generators and to verify oracle-uncompute behavior (\cref{sec:experiments}).
\end{remark}

\begin{remark}[Complete vs incomplete specification]
An ``incomplete'' specification that tests only $\ket{a}\ket{0} \to \ket{a}\ket{x_a}$ leaves half the Hilbert space unconstrained, admitting a continuous family of solutions with many free parameters.
The complete XOR specification---testing $\ket{a}\ket{y}$ for all $y\in\{0,1\}^K$---yields the unique permutation structure in \cref{eq:uqram-product}.
\end{remark}

% ==============================================================================
\section{Relationship to prior paradigms: bucket-brigade and QROM}
\label{sec:prior}
% ==============================================================================

\paragraph{Bucket-brigade \qRAM.}
Routing-based architectures such as the bucket-brigade \qRAM \cite{giovannetti2008qram} implement coherent lookup by dynamically configuring a routing fabric (switches/routers) and sending a bus system to the selected leaf.
At the level of the \emph{entire physical device Hilbert space} (address, routers, bus, and memory degrees of freedom), bucket-brigade corresponds to a single global unitary.
The difference is \emph{mechanism}: routing-based designs rely on sequential path-setup, traversal, and uncomputation, whereas U-\QRAM expresses lookup as direct conditional action with the database register as quantum control.
Noise-robustness and fault-tolerant overhead in routing architectures have been studied extensively \cite{hann2021prxq,dimatteo2020tqe,xu2023micro}.

\paragraph{QROM / lookup tables.}
Circuit-based QROM (quantum read-only memory) and lookup-table techniques (common in fault-tolerant resource estimation) implement XOR-lookup using multi-controlled gates, often with data encoded as \emph{classical} compile-time structure or classical enable signals \cite{babbush2018prx,dimatteo2020tqe}.
U-\QRAM shares the same activation logic (address match \& data bit set) but treats the database as a physical quantum register $M$, so the same unitary acts for all memory configurations.
This distinction matters when the memory must be prepared or updated coherently within a larger quantum computation.

\begin{table}[t]
\centering
\small
\begin{tabular}{@{}lp{3.5cm}p{3.4cm}p{4.0cm}@{}}
\toprule
\textbf{Aspect} & \textbf{Bucket-brigade} & \textbf{QROM} & \textbf{U-\QRAM (this work)}\\
\midrule
Database repr.\ & physical memory & classical parameter & physical register $M$ \\
Global operator & single device unitary & family indexed by data & single unitary on $A{\otimes}D{\otimes}M$\\
Mechanism & routing / bus traversal & compiled gates/pulses & conditional XOR, $M$ as control\\
Quantum data & possible (arch.-dep.) & no & basis-encoded classical (default)\\
Bottleneck & routing errors/latency & compilation overhead & locality/parallelism assumptions\\
\bottomrule
\end{tabular}
\caption{High-level comparison. ``Global operator'' for bucket-brigade refers to the full device Hilbert space including routers/bus; for QROM it is typically a family $\{U_x\}$ as the classical data $x$ changes.}
\label{tab:compare}
\end{table}

\paragraph{Novelty positioning (conservative).}
The novelty here is not the abstract lookup specification \cref{eq:qram-spec}---that is standard.
Nor is any single piece new in isolation: controlled-gate exponentials, orthogonal projector commutation, and one-hot addressing all appear in prior work.
Rather, the contribution is an explicit synthesis of:
\begin{enumerate}[nosep, leftmargin=1.2cm]
\item[(i)] a canonical universal unitary definition \cref{eq:uqram-product} with the database as quantum control (not classical compile-time parameters, not routing-accessed storage),
\item[(ii)] an \emph{exact} commuting-projector Hamiltonian realization (\cref{sec:ham}) that turns ``ballistic \qRAM must be $e^{-iHt}$'' from a critique into a constructive target, and
\item[(iii)] a mode-addressed reduction to uniform 3-local cell terms (\cref{sec:mode}) that sharpens the locality question.
\end{enumerate}
This reframes ``\qRAM practicality'' into a concrete, testable target: can the commuting Hamiltonian be realized with stable phases and non-vanishing couplings as $N$ grows?

% ==============================================================================
\section{Exact commuting-projector Hamiltonian form (no conditional-phase ambiguity)}
\label{sec:ham}
% ==============================================================================

A recurrent pitfall in ``Hamiltonian CNOT'' discussions is the introduction of control-dependent relative phases.
Here we use an exact identity that yields $X$ \emph{exactly} (not merely up to a phase), enabling a clean Hamiltonian form for each controlled-$X$ factor.

\subsection{Exact \texorpdfstring{$X$}{X} from evolution under \texorpdfstring{$(\Id-X)$}{(I-X)}}

\begin{proposition}[Exact $X$ from $\exp(-i\tfrac{\pi}{2}(\Id-X))$]
\label{prop:exactX}
Let $G:=\tfrac{\pi}{2}(\Id-X)$ on a single qubit.
Then $e^{-iG}=X$.
\end{proposition}

\begin{proof}
$X$ has eigenvalues $\pm1$.
Hence $(\Id-X)$ has eigenvalues $0$ and $2$.
Therefore $e^{-i\frac{\pi}{2}(\Id-X)}$ has eigenvalues $1$ and $e^{-i\pi}=-1$, matching the spectrum of $X$ on the same eigenbasis, so it equals $X$.
\end{proof}

\begin{corollary}[Projector-controlled $X$ as an exact exponential]
\label{cor:ctrlXexp}
For any projector $P$ acting on a control subsystem,
\begin{equation}
(\Id-P)\otimes \Id\; +\; P\otimes X
= \exp\!\left(-i\, P\otimes \tfrac{\pi}{2}(\Id-X)\right)
\quad\text{exactly.}
\label{eq:ctrlX-projector}
\end{equation}
\end{corollary}

\subsection{U-\QRAM as one-shot evolution under a commuting sum}

Define, for each cell $(a,j)$,
\begin{equation}
\Hcell
:= (P_a Q_{a,j}) \otimes \frac{\pi}{2}\bigl(\Id - X_{d_j}\bigr).
\label{eq:Hcell}
\end{equation}
By \cref{eq:ctrlX-projector}, $e^{-i\Hcell}$ is exactly the $(a,j)$ controlled-$X$ factor.

\begin{lemma}[Commutation]
\label{lem:commute}
For any two cells $(a,j)$ and $(b,\ell)$,
\begin{equation}
[\Hcell,\ H_{b,\ell}] = 0.
\end{equation}
\end{lemma}

\begin{proof}
All $P_a$ commute and satisfy $P_aP_b=0$ for $a\neq b$.
All $Q_{a,j}$ commute because they act on distinct memory qubits.
If $j\neq \ell$, then $(\Id-X_{d_j})$ and $(\Id-X_{d_\ell})$ act on distinct output qubits and commute.
Thus every factor commutes.
\end{proof}

\begin{theorem}[Commuting-sum Hamiltonian representation (exact)]
\label{thm:commuting-sum}
Let
\begin{equation}
\Htot := \sum_{a=0}^{N-1}\sum_{j=0}^{K-1} \Hcell.
\end{equation}
Then
\begin{equation}
\Uqram
= \exp(-i\Htot)
= \prod_{a=0}^{N-1}\prod_{j=0}^{K-1} \exp(-i\Hcell),
\label{eq:uqram-exp}
\end{equation}
and the equality is exact (no auxiliary phase corrections).
\end{theorem}

\begin{proof}
By \cref{cor:ctrlXexp,eq:Hcell}, each factor of \cref{eq:uqram-product} equals $\exp(-i\Hcell)$.
By \cref{lem:commute}, the $\Hcell$ commute, hence the exponential of the sum equals the product.
\end{proof}

\paragraph{Binary-address locality barrier.}
Under binary addressing, each address projector has the Pauli-product form
\begin{equation}
P_a
= \bigotimes_{t=0}^{n-1} \frac{1}{2}\bigl(\Id + (-1)^{a_t} Z_{A_t}\bigr),
\label{eq:binary-projector}
\end{equation}
so $\Hcell$ is inherently $(n+2)$-local before any gadgetization into 2-local hardware.
The next section identifies an architectural pathway that removes this locality barrier at the Hamiltonian level.

\begin{remark}[Why commutation matters physically]
The commutation structure means that \emph{relative timing and ordering do not affect correctness}.
In a physical implementation, this supports an ``always-on'' interpretation: if all cell couplings can be present simultaneously, the correct lookup emerges from the total evolution.
This is the precise sense in which U-\QRAM admits a single-Hamiltonian query \emph{target}.
\end{remark}

% ==============================================================================
\section{Mode-addressed (one-hot) architecture: uniform 3-local cell terms}
\label{sec:mode}
% ==============================================================================

\subsection{Unary/one-hot addressing}

Instead of encoding $a$ in $n=\log_2N$ qubits, define a unary/one-hot address register $A^{(1)}=(A_0,\dots,A_{N-1})$ constrained to Hamming weight one:
\begin{equation}
\ket{e_a} := \ket{0\cdots 010\cdots 0},
\end{equation}
with the $1$ at position $a$.
A coherent address is a superposition $\sum_a \alpha_a\ket{e_a}$.
In bosonic or photonic encodings, this corresponds naturally to a \emph{single excitation over $N$ modes}.

On the single-excitation subspace, the address projector becomes a single-mode occupation projector,
\begin{equation}
P_a\ \longrightarrow\ \Pone{A_a}
\qquad\text{(or equivalently a number operator $n_a$ on mode $a$).}
\end{equation}

\subsection{3-local Hamiltonian terms}

Substituting $P_a\mapsto \Pone{A_a}$ into \cref{eq:Hcell} yields
\begin{equation}
\Hcellone
:= \bigl(\Pone{A_a}\,Q_{a,j}\bigr)\otimes \frac{\pi}{2}\bigl(\Id - X_{d_j}\bigr),
\label{eq:Hcell-onehot}
\end{equation}
which is \emph{3-local} for every $(a,j)$: (address mode $A_a$) $\otimes$ (memory qubit $m_{a,j}$) $\otimes$ (output qubit $d_j$).

\begin{theorem}[Mode-addressed commuting 3-local Hamiltonian for U-\QRAM]
\label{thm:3local}
Under unary/one-hot addressing,
\begin{equation}
\Uqram = \exp\!\left(-i\sum_{a=0}^{N-1}\sum_{j=0}^{K-1} \Hcellone\right),
\end{equation}
all terms $\{\Hcellone\}$ mutually commute, and each summand is 3-local.
\end{theorem}

\begin{proof}
Identical to \cref{thm:commuting-sum}, replacing $P_a$ by $\Pone{A_a}$ and observing that these are commuting single-qubit projectors across distinct address modes.
\end{proof}

\begin{remark}[Resource trade-off]
Unary addressing exchanges address size ($\log N$ qubits) for locality (from $(\log N+2)$-body terms to 3-local terms).
This trade-off is favorable only in architectures where a mode basis and modal selectivity are natural, and where large parallel couplings are physically plausible.
\end{remark}

\begin{remark}[Single-excitation constraint]
The mode-addressed analysis assumes the single-excitation (Hamming-weight-one) subspace.
Preventing leakage to the zero- or multi-excitation sectors is architecture dependent and must be engineered.
\end{remark}

\subsection{Conceptual architecture}

\begin{figure}[t]
\centering
\begin{tikzpicture}[font=\small, node distance=10mm]
\node[draw, rounded corners, minimum width=10.5cm, minimum height=1.0cm] (addr) {Address (one-hot / modes): $\{\ket{e_a}\}_{a=0}^{N-1}$};
\node[draw, rounded corners, minimum width=10.5cm, minimum height=1.0cm, below=of addr] (mem) {Memory array: qubits $m_{a,j}$ for $a\in[N],\ j\in[K]$};
\node[draw, rounded corners, minimum width=10.5cm, minimum height=1.0cm, below=of mem] (out) {Output register: qubits $d_0,\dots,d_{K-1}$};

\node[draw, rounded corners, fill=gray!10, right=10mm of mem, minimum width=4.0cm, minimum height=2.6cm] (cell) {
\begin{minipage}{3.7cm}
\centering
\textbf{Cell $(a,j)$}\\[1mm]
$\Hcellone=\Pone{A_a}Q_{a,j}{\otimes} \tfrac{\pi}{2}(\Id{-}X_{d_j})$\\[1mm]
(commuting, 3-local)
\end{minipage}
};

\draw[-{Stealth[length=2.0mm]}] (addr.east) -- ++(7mm,0) |- (cell.west);
\draw[-{Stealth[length=2.0mm]}] (mem.east) -- ++(7mm,0) |- (cell.west);
\draw[-{Stealth[length=2.0mm]}] (cell.west) |- (out.east);
\end{tikzpicture}
\caption{Mode-addressed U-\QRAM as a commuting sum of uniform 3-local cell interactions, executable in parallel under explicit assumptions (\cref{sec:latency}).}
\label{fig:architecture}
\end{figure}
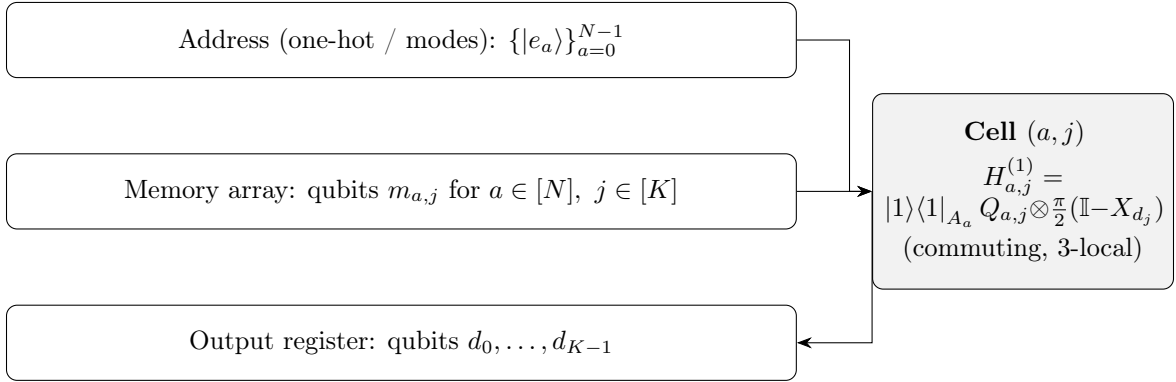

% ==============================================================================
\section{Latency, work, and physical assumptions}
\label{sec:latency}
% ==============================================================================

To avoid equivocation, we separate three notions:
\begin{itemize}[nosep]
\item \textbf{Latency:} wall-clock time for one lookup (what determines end-to-end algorithm runtime).
\item \textbf{Work:} number/extent of physical interactions required (energy, control channels, calibration burden).
\item \textbf{Hardware parallelism:} how many interactions can occur simultaneously.
\end{itemize}
Classical RAM attains low latency via extensive parallel hardware (decoders, wires, sense amplifiers).
Analogously, ``$O(1)$ \qRAM latency'' is meaningful only if one permits either $O(N)$ parallel couplings or a genuinely global bus, and if geometry permits constant-time interaction.

\subsection{Minimal assumptions for constant-latency interpretations}

The commuting-sum form becomes an operational ``one-shot query'' only under assumptions such as:
\begin{quote}
\textbf{A (mode selectivity).} The address is carried by an excitation over $N$ modes so interactions can be conditioned on occupancy of mode $a$ (a projector $\Pone{A_a}$ or number operator $n_a$).

\textbf{B (local controlled interaction).} Each cell $(a,j)$ supports a conditional interaction implementing the 3-local term in \cref{eq:Hcell-onehot} (or an equivalent commuting generator).

\textbf{C (parallelism / global drive).} All cell interactions can occur concurrently (e.g., a single global pulse, a wavefront passing all cells within constant time, or a shared bus coupling to all cells).

\textbf{D (error budget / phase stability).} Address-dependent phases, crosstalk, leakage, and stochastic noise remain within the error budget of the surrounding algorithm.
\end{quote}

\subsection{Pauli expansion and an engineering target}

Writing $\Pone{q}=\tfrac{1}{2}(\Id - Z_q)$, each 3-local term expands into a constant number of commuting Pauli strings.
In particular, the interaction contains components proportional to
\begin{equation}
Z_{A_a} Z_{m_{a,j}} X_{d_j},
\end{equation}
along with one- and two-body terms that, in principle, can be handled via calibration/echo techniques.
A useful engineering target is therefore:
\begin{quote}
Can the platform realize an effective Hamiltonian containing
$\sum_{a,j} g_{a,j}\, Z_{A_a}Z_{m_{a,j}}X_{d_j}$ with \emph{non-vanishing effective strengths} $g_{a,j}$ as $N$ grows,
without uncontrolled crosstalk or address-dependent phase drift?
\end{quote}

\subsection{Causality and geometry}

Even for commuting interactions, finite signal speed and layout can impose lower bounds on physical time.
Fundamental causal bounds for \qRAM architectures formalize this constraint \cite{wang2024causalbounds}.
Our formulation is consistent with such results: constant-latency interpretations require effectively global coupling or colocalized mode interactions (e.g., a cavity bus or integrated photonic setting), and cannot be inferred from commutation alone.

\subsection{Regimes summary}

\begin{table}[t]
\centering
\small
\begin{tabular}{@{}p{2.4cm}p{4.4cm}p{6.0cm}@{}}
\toprule
\textbf{Regime} & \textbf{Model} & \textbf{Scaling takeaway} \\
\midrule
I: Compiled gates & Binary address, standard 2-local gates & Address decoding reintroduces $\Omega(\log N)$ depth; compiled cost depends on ancilla trade-offs. \\
\addlinespace
II: Special primitives & Global fanout-like gates, nonstandard multi-qubit primitives & Depth can be reduced (sometimes to constant) under strong hardware assumptions. \\
\addlinespace
III: Direct $\exp(-i\Htot)$ & Engineered commuting interaction (global bus / parallel couplings) & Constant latency requires non-vanishing couplings as $N$ grows and geometry permitting constant-time interaction. \\
\bottomrule
\end{tabular}
\caption{Latency regimes: the main contribution is an explicit commuting-Hamiltonian target; constant-latency interpretations live in Regime III under stated assumptions.}
\label{tab:regimes}
\end{table}

% ==============================================================================
\section{Compiled circuit baselines, exact counts, and resource accounting}
\label{sec:compiled}
% ==============================================================================

\subsection{Exact gate count at the logical level}

For binary addressing, \cref{eq:uqram-product} corresponds directly to $NK$ multi-controlled $X$ operations: one per cell $(a,j)$.
Each such operation has $(\log_2 N + 1)$ controls (address-equality plus one memory control) and one target $d_j$.
These factors commute (\cref{lem:factors-commute}), so at the logical level they may be sequenced in any order, and in principle parallelized across $a$ if the hardware supports it.

\paragraph{Complexity comparison.}
The gate complexity of U-\QRAM is $O(NK)$ multi-controlled gates, each with $O(\log N)$ controls.
This is dramatically more efficient than a generic $2^n \times 2^n$ unitary decomposition ($O(4^n)$ gates) and comparable to bucket-brigade node counts while avoiding sequential routing traversal.
For $N=4$, $K=1$ (a $128 \times 128$ unitary matrix): generic decomposition requires $O(16{,}000)$ gates; U-\QRAM requires exactly 4 multi-controlled gates.

\subsection{2-qubit compilation scaling}

When restricted to a standard 2-qubit gate set, multi-controlled $X$ gates must be decomposed.
A conservative, hardware-agnostic scaling upper bound (up to ancilla/depth trade-offs and improved constants) is
\begin{equation}
\mathrm{Depth}_{\mathrm{active}} = \Theta(NK\log N),
\qquad
\#(2\text{-qubit gates}) = \Theta(NK\log N),
\end{equation}
using standard decompositions of multi-controlled Toffoli/CNOT into 2-qubit gates \cite{selinger2013tdepth1,jones2013toffoli}.
Fault-tolerant resource estimation can substantially amplify these costs \cite{dimatteo2020tqe}.

\begin{table}[t]
\centering
\small
\begin{tabular}{@{}lcc@{}}
\toprule
\textbf{Quantity} & \textbf{Binary address} & \textbf{Unary (one-hot) address} \\
\midrule
Address size & $\log_2 N$ qubits & $N$ modes (single-excitation) \\
Memory size & $NK$ qubits & $NK$ qubits \\
Output size & $K$ qubits & $K$ qubits \\
Cell term locality & $(\log N+2)$-local & 3-local \\
Number of cell terms & $NK$ & $NK$ \\
\bottomrule
\end{tabular}
\caption{Resource and locality trade-offs. Unary addressing increases address size but reduces cell locality.}
\label{tab:resources}
\end{table}

\paragraph{Interpretation.}
In Regime I (compiled 2-local gates), U-\QRAM is not claimed to outperform conventional QROM constructions on latency.
Its value is structural: it provides a single data-independent unitary, an exact commuting-Hamiltonian target, and an explicit architectural condition (mode addressing) under which constant-locality interactions are possible.

% ==============================================================================
\section{Algorithmic composability: Grover-oracle usage}
\label{sec:grover}
% ==============================================================================

A standard pattern in algorithms such as Grover search is to compute a predicate on the retrieved data and then uncompute the lookup:
\begin{equation}
U_{\mathrm{oracle}} = \Uqram\, U_{\mathrm{check}}\, \Uqram,
\end{equation}
where $U_{\mathrm{check}}$ applies a phase (or a flag) conditioned on the data register $D$.

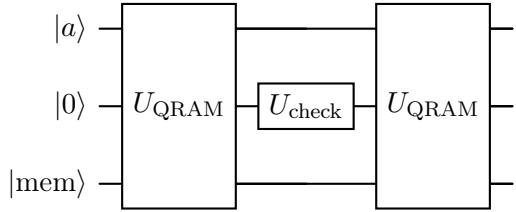
\begin{figure}[ht]
\centering
\begin{quantikz}[row sep=0.4cm, column sep=0.3cm]
\lstick{$\ket{a}$} & \gate[3]{\Uqram} & \qw & \gate[3]{\Uqram} & \qw \\
\lstick{$\ket{0}$} & & \gate{U_{\text{check}}} & & \qw \\
\lstick{$\ket{\text{mem}}$} & & \qw & & \qw
\end{quantikz}
\caption{Grover oracle using U-QRAM. The forward lookup retrieves data, $U_{\text{check}}$ marks matches with a phase flip, and the second $\Uqram$ (self-inverse) uncomputes the retrieval.}
\label{fig:grover-oracle}
\end{figure}

Because $\Uqram$ is self-inverse (\cref{cor:selfinverse}), the lookup-unlookup wrapper is straightforward and makes phase-stability requirements explicit.
In mode-addressed proposals, a central requirement is that this sandwich does \emph{not} introduce address-dependent phases that would spoil Grover interference (\cref{sec:experiments}).

\paragraph{Integration advantages.}
Because U-\QRAM is a standard unitary composed of multi-controlled gates:
\begin{itemize}[nosep]
\item \textbf{No hybrid control:} The entire computation---including data loading---is a coherent quantum circuit with no classical intervention mid-execution.
\item \textbf{Trivial uncomputation:} The self-inverse property means uncomputation requires no additional circuit design.
\item \textbf{Composability:} Memory can be prepared by preceding quantum operations, enabling pipelines where one algorithm's output becomes another's data.
\end{itemize}
These properties are particularly valuable for quantum machine learning, where algorithms typically require repeated data loading and uncomputation within a single coherent execution.

% ==============================================================================
\section{Experimental path forward}
\label{sec:experiments}
% ==============================================================================

To ground the theoretical construction in physical reality, we identify concrete experimental milestones that would validate or refute the mode-addressed commuting-Hamiltonian thesis.

\paragraph{Milestones.}
Three experimentally meaningful (and refutable) steps directly test the mode-addressed commuting-Hamiltonian thesis:
\begin{enumerate}[leftmargin=1.15cm, itemsep=2pt]
\item \textbf{Mode-addressed demonstration at modest scale.}
Implement $N\in\{8,16\}$ with $K=1$ where addresses are one-hot/mode superpositions, memory bits are physical qubits, and a single global evolution approximates $\Uqram$.
\item \textbf{Phase stability for oracle uncompute.}
Demonstrate Grover-style usage $U_{\mathrm{oracle}}=\Uqram\,U_{\mathrm{check}}\,\Uqram$ without address-dependent phase drift.
\item \textbf{Scaling study of effective interactions and crosstalk.}
Measure how effective conditional couplings and crosstalk behave as the number of simultaneously coupled cells increases, confronting causal and architectural constraints \cite{wang2024causalbounds}.
\end{enumerate}

\paragraph{Candidate hardware directions.}
Mode-addressing and multi-mode coupling arise naturally in integrated photonics and in multimode superconducting platforms (microwave resonators or hybrid acoustic/phononic systems).
The role of this paper is not to select a winner, but to provide a clean Hamiltonian target (\cref{eq:Hcell-onehot}) that such platforms can aim to approximate and a clear checklist for what must scale.

% ==============================================================================
\section{Conclusion}
% ==============================================================================

We presented U-\QRAM as a universal lookup unitary in which the database is a physical memory register acting as quantum control.
The key insights are:
\begin{enumerate}[nosep]
\item Memory qubits act as control signals, not passive storage to be routed to.
\item The unitary has a natural block-diagonal permutation structure, unique under the complete XOR specification.
\item An exact commuting-projector Hamiltonian representation $\Uqram=\exp(-i\Htot)$ provides a precise ``single-Hamiltonian query'' target without conditional-phase ambiguity.
\item Under unary (one-hot, mode-addressed) encoding, each cell interaction becomes uniform and 3-local.
\end{enumerate}

The contribution is not any single piece---controlled-gate exponentials, commuting projectors, and one-hot addressing all exist in prior work---but the explicit synthesis into a canonical target with stated assumptions.
The central open question is whether the commuting Hamiltonian can be realized with stable phases, controlled leakage/crosstalk, and non-vanishing effective couplings as $N$ grows.
By making assumptions and targets explicit, U-\QRAM reframes \qRAM practicality into a concrete program rather than an oracle assumption.

% ==============================================================================
\appendix
\section{Small-\texorpdfstring{$N$}{N} circuit illustrations}
\label{app:circuits}

The projector-form definition \cref{eq:uqram-product} corresponds directly to $NK$ multi-controlled $X$ gates.
For intuition, we include two small cases.

\subsection{Case \texorpdfstring{$N=2$, $K=1$}{N=2, K=1}}

\begin{figure}[H]
\centering
\begin{quantikz}[row sep=0.35cm, column sep=0.45cm]
\lstick{$\ket{a_0}$} & \octrl{1} & \ctrl{1} & \qw \\
\lstick{$\ket{d_0}$} & \targ{}  & \targ{}  & \qw \\
\lstick{$\ket{m_{0,0}}$} & \ctrl{-1} & \qw & \qw \\
\lstick{$\ket{m_{1,0}}$} & \qw & \ctrl{-2} & \qw
\end{quantikz}
\caption{U-\QRAM for $N=2$, $K=1$: flip $d_0$ iff $a=0$ and $m_{0,0}=1$, and also iff $a=1$ and $m_{1,0}=1$.}
\label{fig:n2k1}
\end{figure}
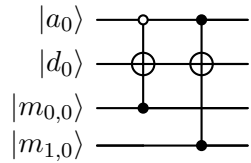

\subsection{Case \texorpdfstring{$N=4$, $K=1$}{N=4, K=1}}

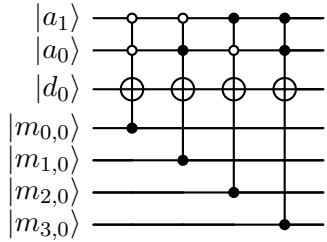
\begin{figure}[H]
\centering
\begin{quantikz}[row sep=0.28cm, column sep=0.35cm]
\lstick{$\ket{a_1}$} & \octrl{1} & \octrl{1} & \ctrl{1} & \ctrl{1} & \qw \\
\lstick{$\ket{a_0}$} & \octrl{1} & \ctrl{1} & \octrl{1} & \ctrl{1} & \qw \\
\lstick{$\ket{d_0}$} & \targ{}  & \targ{}  & \targ{}  & \targ{}  & \qw \\
\lstick{$\ket{m_{0,0}}$} & \ctrl{-1} & \qw & \qw & \qw & \qw \\
\lstick{$\ket{m_{1,0}}$} & \qw & \ctrl{-2} & \qw & \qw & \qw \\
\lstick{$\ket{m_{2,0}}$} & \qw & \qw & \ctrl{-3} & \qw & \qw \\
\lstick{$\ket{m_{3,0}}$} & \qw & \qw & \qw & \ctrl{-4} & \qw
\end{quantikz}
\caption{U-\QRAM for $N=4$, $K=1$: four controlled-$X$ operations, each enabled by one address pattern and one memory bit.}
\label{fig:n4k1}
\end{figure}

% ==============================================================================
\section*{Acknowledgments}
% ==============================================================================

The author thanks Steven M.\ Girvin for helpful correspondence and detailed feedback that sharpened the distinction between U-\QRAM and QROM architectures.

% ==============================================================================
% Bibliography
% ==============================================================================

\end{document}